# Discovery of room temperature multiferroicity and magneto-electric coupling in $Fe_3Se_4$ nanorods

Mousumi Sen Bishwas, and Pankaj Poddar*

## Abstract

We report for the first time, that $Fe_3Se_4$ is a room temperature, type-II multiferroic with magnetoelectric coupling. We observed the coexistence of coupled ferrimagnetic and ferroelectric ordering in $Fe_3Se_4$nanorods well above room temperature, which is a hard magnet with large magnetocrystalline anisotropy. For the first time, we observed spontaneous, reversible ferroelectric polarization in $Fe_3Se_4$ nanorods below the magnetic Curie temperature. The coupling is manifested by an anomaly in the dielectric constant and Raman shift at $T_c$. We do not completely understand the origin of the ferroelectric ordering at this point however the simultaneous presence of magnetic and ferroelectric ordering at room temperature in $Fe_3Se_4$ along with hard magnetic properties will open new research areas for devices.

* Email: p.poddar@ncl.res.in

**Introduction:**

Since the renaissance of multiferroics in previous decade[1]the search for single-phase multiferroics with room temperature magneto-electric coupling has not yielded much result. There are, in fact very few materials discovered so far that show the coexistence of ferromagnetic (FM) ordering along with ferroelectric (FE)one, though some compounds like $BiFeO_3$ became hugely popular due to their antiferromagnetic (AFM)property along with properties[2–4]. Some of the non-stoichiometric compounds have achieved some of the desired propertiesthrough doping [5–7], although they are hard to reproduce and not suitable for large scale production. Some of the technical challenges associated with most of the compounds are very low magnetic transition temperature, small polarization values and week coupling between them etc. However, multiferroic compounds are full of surprises. Lately, there were reports for unconventional multiferroics where mechanisms for the origin of ferroelectricity were discussed in centro-symmetric materials based on charge ordering and bond ordering[8]. $Fe_3O_4$ has become the oldest known multiferroic to human kind after the discovery of spontaneous polarization in $Fe_3O_4$below 38 K arising from alternation of charge states and bond lengths[9–11].

Transition metal based chalcogenides possesses huge potential as magneto-electric materials. $CdCr_2S_4$ and $CdCr_2Se_4$with Curie temperature 85 K and 125 Krespectively were reported to show multiferroic behavior,[12] although the origin of this behavior was not clearly understood. Recently, high temperature multiferroicity was predicted in $BaFe_2Se_3$ phase, which was quite unexpected[13]. Later, it was shown experimentally with the help of photoemission spectroscopy that two kinds of electrons (localized and itinerant) coexist in $BaFe_2Se_3$ phase[14].

Binary transition metal chalcogenide $Fe_3Se_4$ has gained attention quite recently because of its high uniaxial magnetic anisotropy constant without the presence of any rare earth metal or noble metal atom leading to very high coercivity at room temperature (~4kOe)[15–17]. $Fe_3Se_4$ is ferrimagnetic at room temperature with Curie temperature 323 K and have monoclinic structure with a space group I2/m (12) [16]. We have measured and discussed the magnetic entropy and heat capacity of$Fe_3Se_4$ nanorods in our previous work[18].But,till now, dielectric study of this material has not been reported either in bulk or nanoforms.

In this report we investigate a novel and unexpected perspective of $Fe_3Se_4$ which establishes $Fe_3Se_4$ as a room temperature multiferroic material. $Fe_3Se_4$ shows magneto-electric coupling at room temperature along with very strong ferrimagnetic properties which is very unique for single phase binary compound in undoped form with a simple crystal structure.These behaviors are very surprising for $Fe_3Se_4$, as, at room temperature the single crystals of this compound is known to show metallic properties due to the overlapping of cation-cation forming bands[19].Metallic behavior and ferroelectricity are not mutually compatible as the conduction electrons screen the static internal electric field. This kind of dual behavior was predicted verylong ago[20]but found to exist recently in $LiOsO_3$[21].

**Sample preparation:**

In this work, we have used the sample from our previous work mentioned in ref.18. The details of sample synthesis and basic characterizations are mentioned in the supporting information.

**Experimental details and techniques:**

Magnetization measurements were done using the VSM attachment of PPMS from Quantum Design systems equipped with 9 T superconducting magnet on powder samples packed in special plastic holders designed so that the dielectric contribution of the holder is negligible..

Temperature dependent dielectric spectroscopy was performed using Novocontrol Beta NB Impedance Analyzer connected with home built sample holder to couple with a helium closed cycle refrigerator (Janis Inc.). The powdered sample was compressed in the form of circular pellet of diameter 13 mm and a custom designed sample holder was used to form parallel plate capacitor geometry.Ferroelectric hysteresis loop measurements were done on pellets made by cold pressing the sample powder in zero field.Raman spectra were recorded on an HR-800 Raman spectrophotometer (JobinYvon-Horiba, France) using monochromatic radiation emitted by a He−Ne laser(633 nm), operating at 20 mW and with accuracy in the range between 450 and 850 nm ± 1 cm−1. An objective of 50× LD magnification was used both to focus and to collect the signal from the powder sample dispersed on the glass slide. For magneto-Raman measurements, a permanent bar magnet was used to apply magnetic field near the sample. The intensity of magnetic field was measured with Gaussmeter by placing the hall probe as near as possible to the sample.

**Results and discussion:**

Nanorods of $Fe_3Se_4$, grown by high temperature thermal decomposition in organic solvents, with average diameter ~50 nm (supporting information). X-ray diffraction pattern showed that nanorods are monoclinic in structure with space group I2/m (12)(supporting information). The magnetic property of these samples were studied in detail in our previous works[16,18].

**Magnetic property of $Fe_3Se_4$ nanorods:**

Temperature dependence of magnetization in zero field cooled (ZFC) and field cooled conditions (FC) shows bifurcation below 340 K and Curie transition temperature around 323 K, below which it goes into a ferrimagnetic phase. Room temperature (300 K) hysteresis measurements yield a closed loop with coercivity 2.6 kOe and saturation magnetization 5 emu/g. As the temperature is lowered to 10 K, huge increase in coercivity occurs reaching a value of 30 kOe. These values are well according to the values reported in earlier reports in $Fe_3Se_4$ nanoparticles[15–17]. The hard magnetic property in $Fe_3Se_4$ comes from large anisotropy constant which is a result of ordered iron vacancies in alternative layers of iron in lattice[22,23]. Theorigin and behavior of magnetic properties are discussed in detail in our early reports[16,18].

**Observation of electrical polarization in $Fe_3Se_4$ nanorods:**

The as synthesized sample pressed in the form of pellet shows the evidence of the presence of spontaneous and reversible polarization in $Fe_3Se_4$ nanorods. Figure 1 shows the P-E loop taken with frequency 500 Hz and at a potential of 100 V at room temperature (~300 K). No electrical poling of the sample was done before the measurement. The loop displays clear ferroelectric hysteresis. The hysteresis loop is symmetrical and does not show clear saturation. The shape of the loop resembles the hysteresis loop observed for $Fe_3O_4$ nanoparticles and its composites with PVDF[24]. The value of polarization is very small but comparable to other nanoparticles systems[12,25,26]. The maximum polarization achieved for 100 V is 0.012 $\mu C/cm^2$at room temperature.Within the accessible instrumental range of potential (100 V), no real saturation value of polarization is obtained. The polarization can be increased with electrically poling the samples to orient the dipoles prior the hysteresis measurement.

There are several methods to rule out the possibility of getting an artifact in the ferroelectric hysteresis measurements. One of the most straightforward way is to measure the polarization at different frequencies, as artifacts are usually highly frequency dependent. Figure 4.3 shows the polarization hysteresis loop taken with 100 V potential at various frequencies.[27]. A closed loop is observed in all cases spanning a broad frequency region (100-500 Hz). As it is clear from the figure 4.3 the area under the ferroelectric loop increases with decrease in the frequency but the shape of the loop remains same at all frequencies. Therefore, the occurrence of artifacts can be ruled out in this case.

**Impedance spectroscopy:**

The dielectric response from thesample was measured in a frequencyrange 1 Hz to $10^6$ Hz spanning temperature range150 K to 350 K at 1 V rms (see supporting information).The frequency dependence of various parameters is plotted as a function of frequency in FigureS3.As perceived from the figure, both the real and out of phase part of permittivity (ε' and ε" respectively) values decreases with increasing frequency. At low frequency, the permittivity values consists of contributions from all the dipolar, interfacial, atomic, ionic and electronic polarization, which can be explained by Maxwell-Wagner theory. The heavier dipoles are able to follow the external field at low frequency, so that values of ε are higher. As the frequency starts to increase the dipoles lag behind the field and ε value decreases.In many compounds large values of permittivity are seen owing to the presence of ionic impurities in the sample. This is revealed by high values of tan δ in these cases. Here, tan δ values of the order $10^{-1}$ are achieved (see figureS3c). At frequency greater than $10^4$ Hz, a frequency dependent feature is seen in tan δ plot. The origin of this frequency dependence in loss tangent is not clearly understood.

The temperature dependence of ε' is extracted from the above figure (Figure S3)and shown in figure 2 at frequencies 107, 1587 and 11952 Hz.Asmall kink was observed around temperature 323 K in real part of ε verses temperature curve for all frequencies. For clear view the temperature dependence of dielectric constant and loss tangent is plotted in bottom panel of figure 2. The kink around the transition temperature is encircled.Above 317 K, the value of ε decreases sharply indicative of a ferroelectric to paraelectric transition.Although the feature seen in very weak in nature, but the proximity of the kink observed with the magnetic Curie transition (Tc) in this compound indicates towards the presence of magneto-electric coupling in the compound.

To understand the nature of dielectric relaxation in these nanorods, complex Argand plane plot ε'' and ε', also known as Cole-Cole plot, was examined (see supporting information). FigureS4 shows the plots between the real (ε')and imaginary (ε'') part of the impedanceat different temperatures. It can berealized from the Cole-Cole plot that the center of the semicircle arcs are below the x-axisimplying that the electrical response from the sample departs from the ideal Debye's relaxation process. As can be inferred from the figure, at very low temperatures (plots at 185 K, 220 K), there is only one semicircle which comes from the contribution from the grains. As the temperature increases and approaches near the Curie temperature another semicircle appears at lower frequency region associated with the contribution from the grain boundary (mostly the organic surfactants used in the synthesis). At Tc (323 K), both the contributions from grain and grain boundary is prominent but when the temperature is raised much above Tc (plots at 336 K and 350 K), only the contributions due to grain boundary is prominent. The resistance from the grain and grain boundaries are calculated from the Cole-Cole plots and plotted against temperature in figure 3. The parameter α(spreading factor) characterizes the distribution of relaxation time signifying the departure from the ideal electrical response. αcan be determined from the expression for the maximum value of the imaginary part of the permittivity given below.

$$\epsilon''_{max} = \frac{(\epsilon_s - \epsilon_\infty)\tan\left[(1-\alpha)\,{}^{\pi}/_{4}\right]}{2}$$

Here, $\epsilon_s$ and $\epsilon_\infty$are the low and high frequency limit of $\epsilon'$respectively. The spreading factor αwas calculated from the above equation for both the semicircle arcs and the values for grain and grain boundary were found to be 0.54 and 0.57, respectively at 317 K. These non-zero values of αshows the poly-dispersive nature of dielectric relaxation as observed in literatures [28,29].

In order to study the coupling between spin and the phonons in the present system Raman spectrum was measured from temperature 295 to 333 K. Figure 4shows the Raman spectra of as prepared $Fe_3Se_4$ nanoparticles taken at temperaturesfrom 295 K to 333 K. The spectrum consists of sharp peaks at 224, 291, 409 $cm^{-1}$. The peak at 224 and 291 $cm^{-1}$ can be ascribed to the Fe-Se vibration modes as it is close to the reported values of 220 and 285 $cm^{-1}$ for the Fe-Se vibration in β-$Fe_7Se_8$ having similar monoclinic structure[30,31].As the temperature is increased, the peaks, 224, 291 $cm^{-1}$, show significant change in the peak position and peak width (FWHM).

To analyze the spectra, least square fit with Lorentzian line shape was used to fit the peaks. When the peak position and FWHM is plotted against temperature, a clear incongruity is seen around the magnetic/ferroeletric transition temperature (~ 323 K) (see figure6). From the figure4, it is appreciated that both the Raman modes softens as temperature is increased from 295 K. As the temperature further reaches the magnetic/ferroelectric ordering temperature the Raman mode starts hardening and peak shifts towards higher wavenumber and then immediately after $T_c$decreases sharply towardslower wavenumber. This anomaly in Raman modes observed near magnetic $T_c$provided a significant input indicating the presence of spin-phonon coupling in the system. This kind of anomaly near the magnetic transition temperature has been observed previously in case of some rare earth chromites[32], pure selenium element[33]and the concurrent anomaly around $T_c$ is ascribed to the spin phonon coupling.

For $Fe_3Se_4$ single crystals it was observed that beyond magnetic ordering temperature the interatomic spacing rearranges such that the cation-cationoverlapping disappears partially or completely[19].

**Observation of spin-phonon-charge coupling:**

Evidence for the presence of spin-charge-phonon coupling in this system can been seen from figure 5 where the magnetization (M), real part of dielectric permittivity (ε'), specific heat capacity at constant pressure ($C_p$) and the heat flow curves taken from thermo gravimetric analysis (TGA) measurements are plotted as a function of temperature. The anomaly in these parameters are highlighted in the shaded region.

**Conclusion:**

In this work, we report the observation of ferroelectric order in $Fe_3Se_4$ nanoparticles at room temperature. These particles alsoshow signatures of spin-charge coupling as an anomaly was observed in dielectric permittivity around magnetic transition temperature 323 K. Ferroelectric polarization measurements revealed hysteresis loops in a broad frequency range. The microscopic origin of the coupling between spin-charge-phonon is not clearly understood. Vigorous theoretical calculations are required to probe this mechanism in this compound.Our important observation about the coexistence of both magnetic and charge ordering at room temperature proposes $Fe_3Se_4$ as a possible room temperature multiferroic compound.

**Acknowledgement:**

P.P. acknowledges the Centre for Excellence in Surface Science at the CSIR-National Chemical Laboratory, network project on Nano-Safety, Health &Environment (SHE) funded by the Council of Scientific and IndustrialResearch (CSIR), India, and the Department of Science & Technology (DST), India through and Indo-Israel grant to develop materials for solar-voltaic energy devices DST/INT/ISR/P-8/2011. M.S.B acknowledges the support from the DST, Indiafor providing Junior Research Fellowship (SRF) through the INSPIRE program.

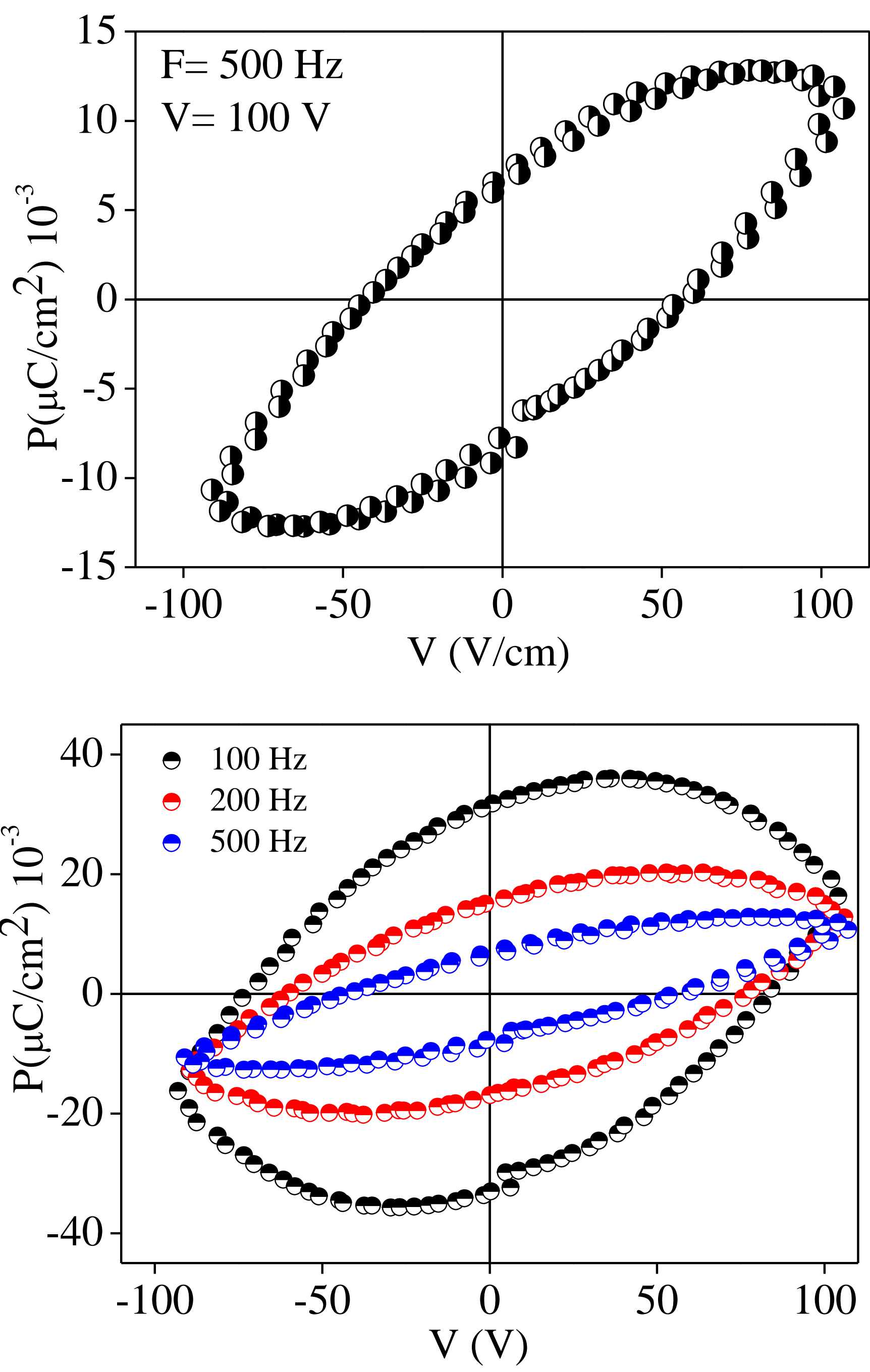

**Figure 6:** Ferroelectric polarization loop of $Fe_3Se_4$nanoparticles at frequency 500 Hz and 100 V applied voltage (top panel). The frequency dependence of the loop taken at frequency 100, 200 and 500 Hz (bottom panel).

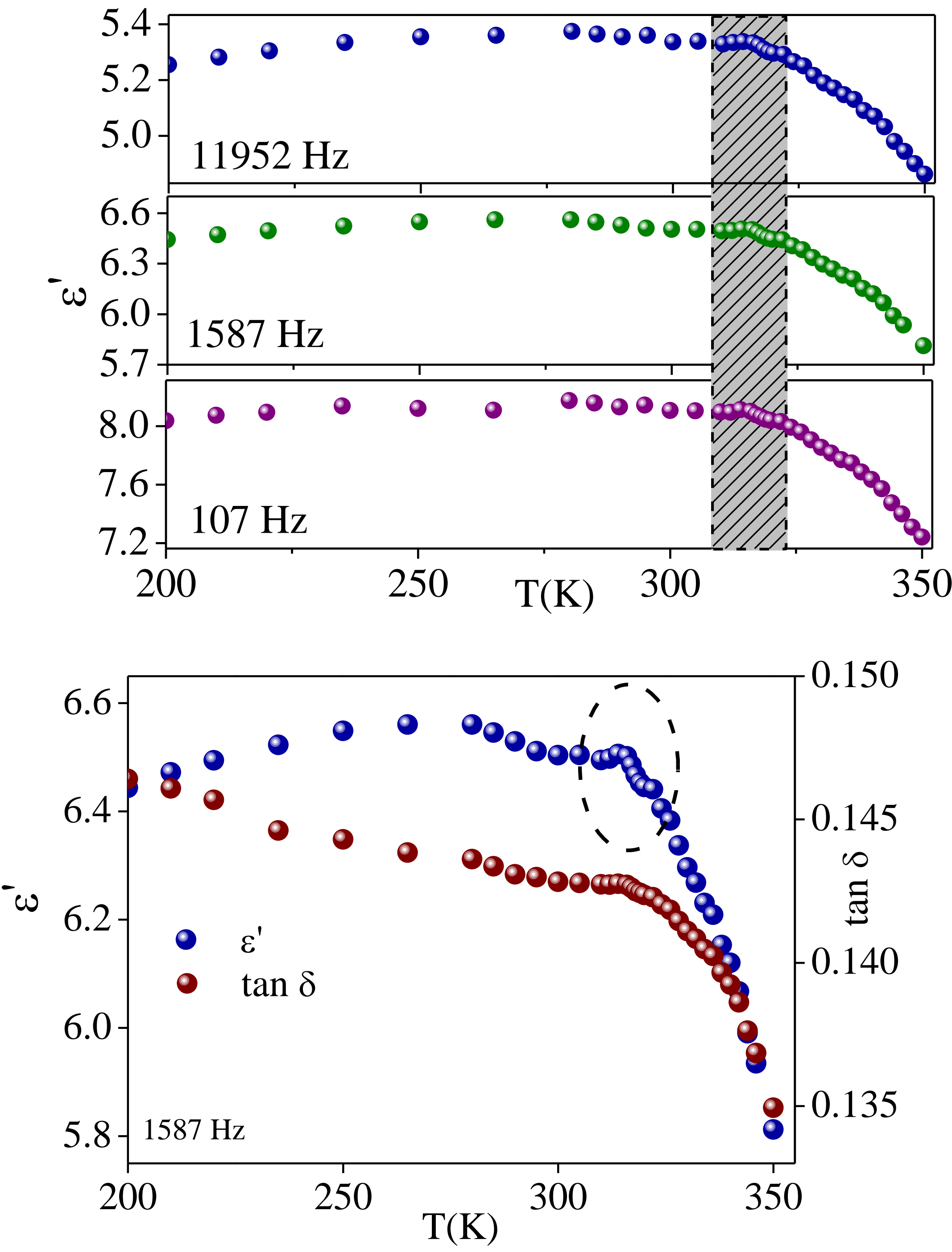

**Figure 2(a).** Temperature dependent plot of real part of permittivity from 200 K to 350 K at frequencies 107 Hz, 1587 Hz and 11952 Hz respectively, extracted from the figure 1 (a). The shaded region highlights the weak anomaly observed

**(b)** Temperature dependence of real part of permittivityand loss tangent at frequency1587 Hz. An anomaly can be seen (encircled) in ε' around 317 K which is in close proximity of magnetic transition temperature.

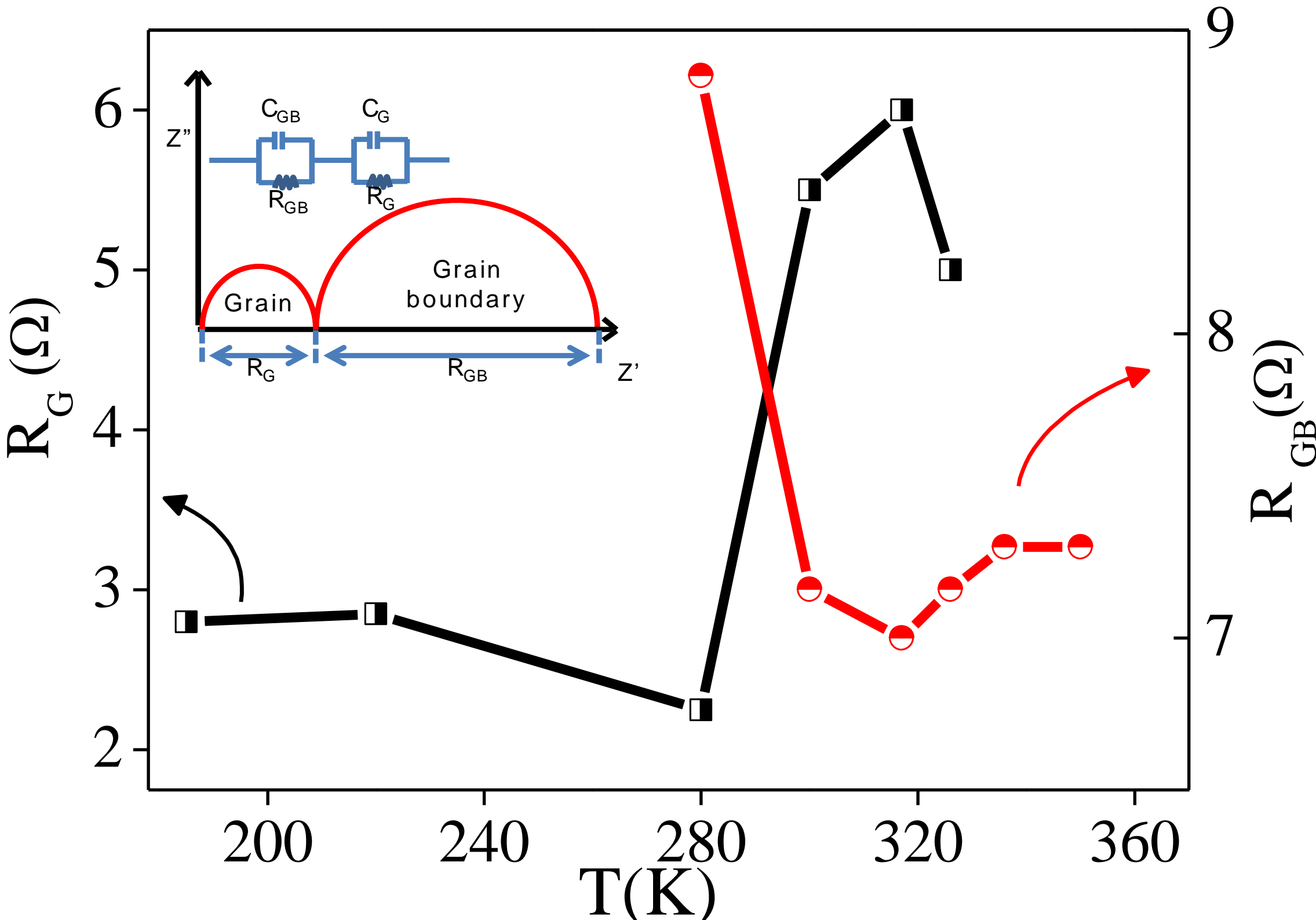

**Figure 3:** Temperature dependent of resistance contribution from grain and grain boundary extracted from Cole-Cole plot is plotted. Inset shows the circuit arrangements and parameters $R_G$ and $R_{GB}$ in a Cole-Cole plot.

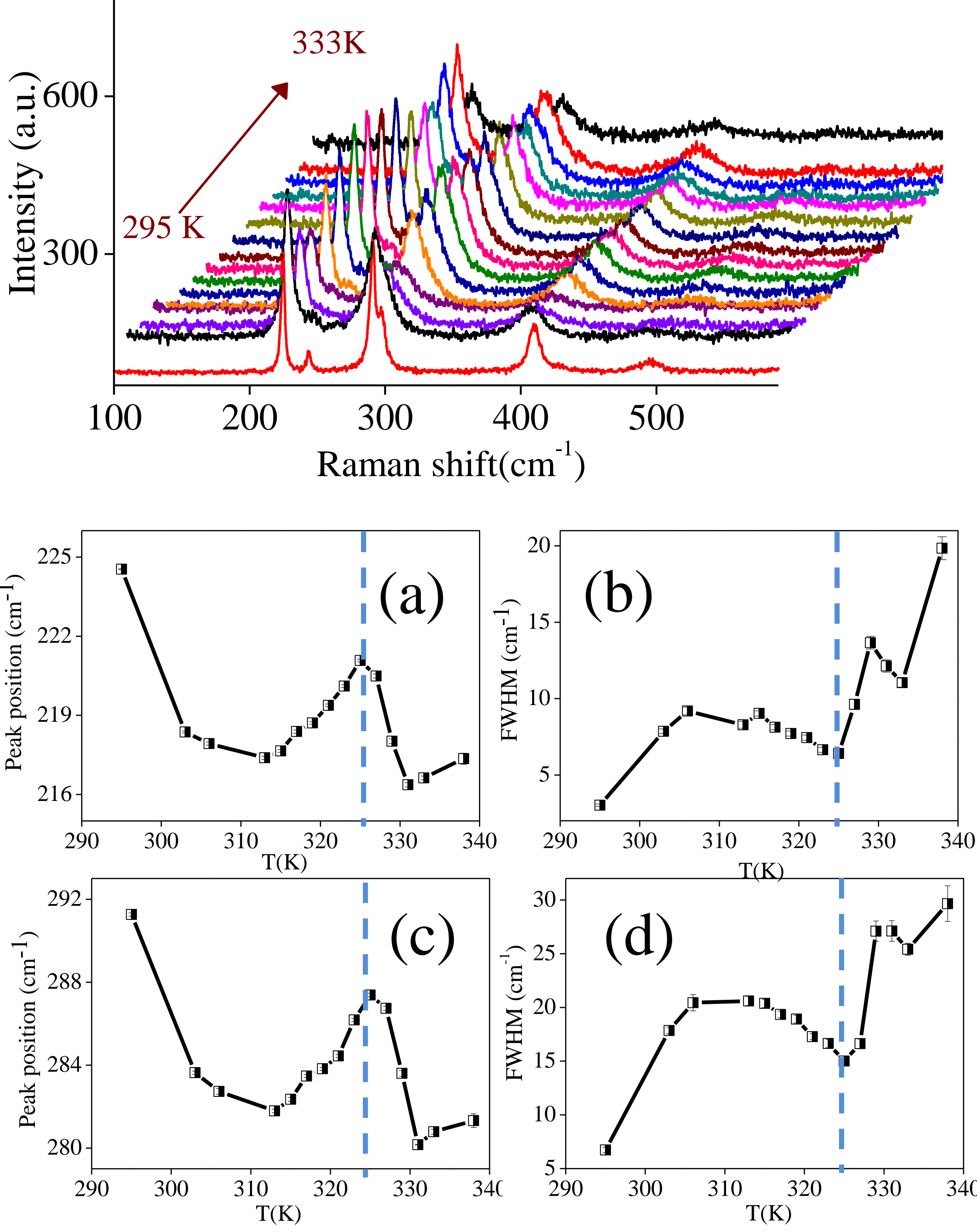

**Figure4**:Temperature dependence of Raman scattered signal from $Fe_3Se_4$ nanoparticles recorded from temperature 295 K to 333 K (top panel).The peak position and FWHM of two Raman modes were deduced from these plots and plotted against temperatures (Bottom panel).

(a)and (b) shows the variation for 224 $cm^{-1}$mode and (c) and (d) shows the variation for 291 $cm^{-1}$mode

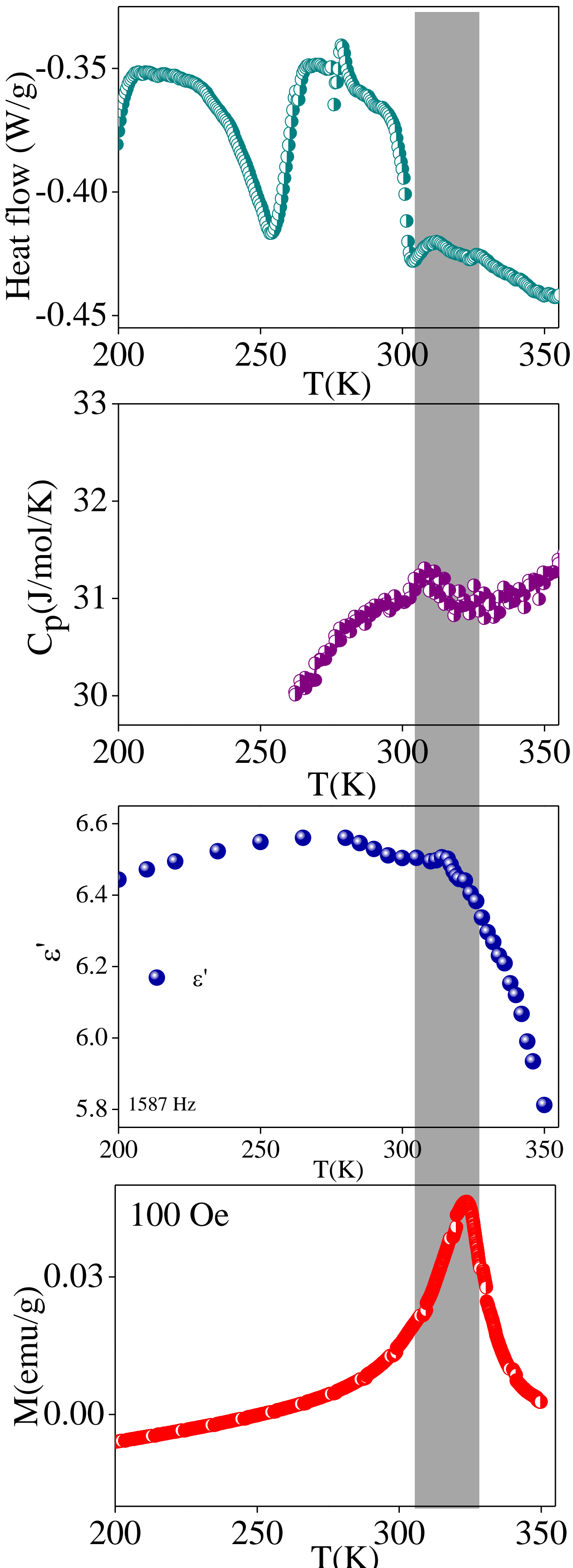

**Figure 5:** Understanding the spin-phonon-charge coupling in $Fe_3Se_4$ at room temperature. The shadowed region shows anomaly in magnetization, dielectric permittivity, heat capacity and heat flow measured from thermo-gravimetric measurements.

## Supporting information

### Synthesis of $Fe_3Se_4$ nanoparticles by one pot organic phase synthesis:

$Fe(acac)_3$ (0.53 g, 1.5 mmol) and Se powder (0.158 g, 2 mmol) were added to 15 ml of oleylamine in a 100 ml three-neck flask under $N_2$ atmosphere. The mixture was heated to 120 °C and kept for 1 h. Then, temperature was increased up to 200 °C and kept for 1 h. Finally, the solution temperature was raised to 300 °C and kept for 1 h. After 1 h, the heat source was removed and solution was allowed to cool down naturally to room temperature. The $Fe_3Se_4$ nanoparticles were precipitated by the addition of 20 ml of 2-propanol. The precipitate was then centrifuged and washed with solution containing hexane and 2-propanol in 3:2 ratio.

### Structural and morphology characterization of $Fe_3Se_4$ nanoparticles:

### Experimental details and techniques:

Magnetization measurements were done using the VSM attachment of PPMS from Quantum Design systems equipped with 9 T superconducting magnet on powder samples packed in special plastic holders designed so that the dielectric contribution of the holder is negligible..

Temperature dependent dielectric spectroscopy was performed using Novocontrol Beta NB Impedance Analyzer connected with home built sample holder to couple with a helium closed cycle refrigerator (Janis Inc.). The powdered sample was compressed in the form of circular pellet of diameter 13 mm and a custom designed sample holder was used to form parallel plate capacitor geometry. Ferroelectric hysteresis loop measurements were done on pellets made by cold pressing the sample powder in zero field. Raman spectra were recorded on an HR-800 Raman spectrophotometer (Jobin Yvon-Horiba, France) using monochromatic radiation emitted by a He−Ne laser (633 nm), operating at 20 mW and with accuracy in the range between 450 and 850 nm ± 1 cm−1,. An objective of 50× LD magnification was used both to focus and to collect the signal from the powder sample dispersed on the glass slide. For magneto-Raman measurements, a permanent bar magnet was used

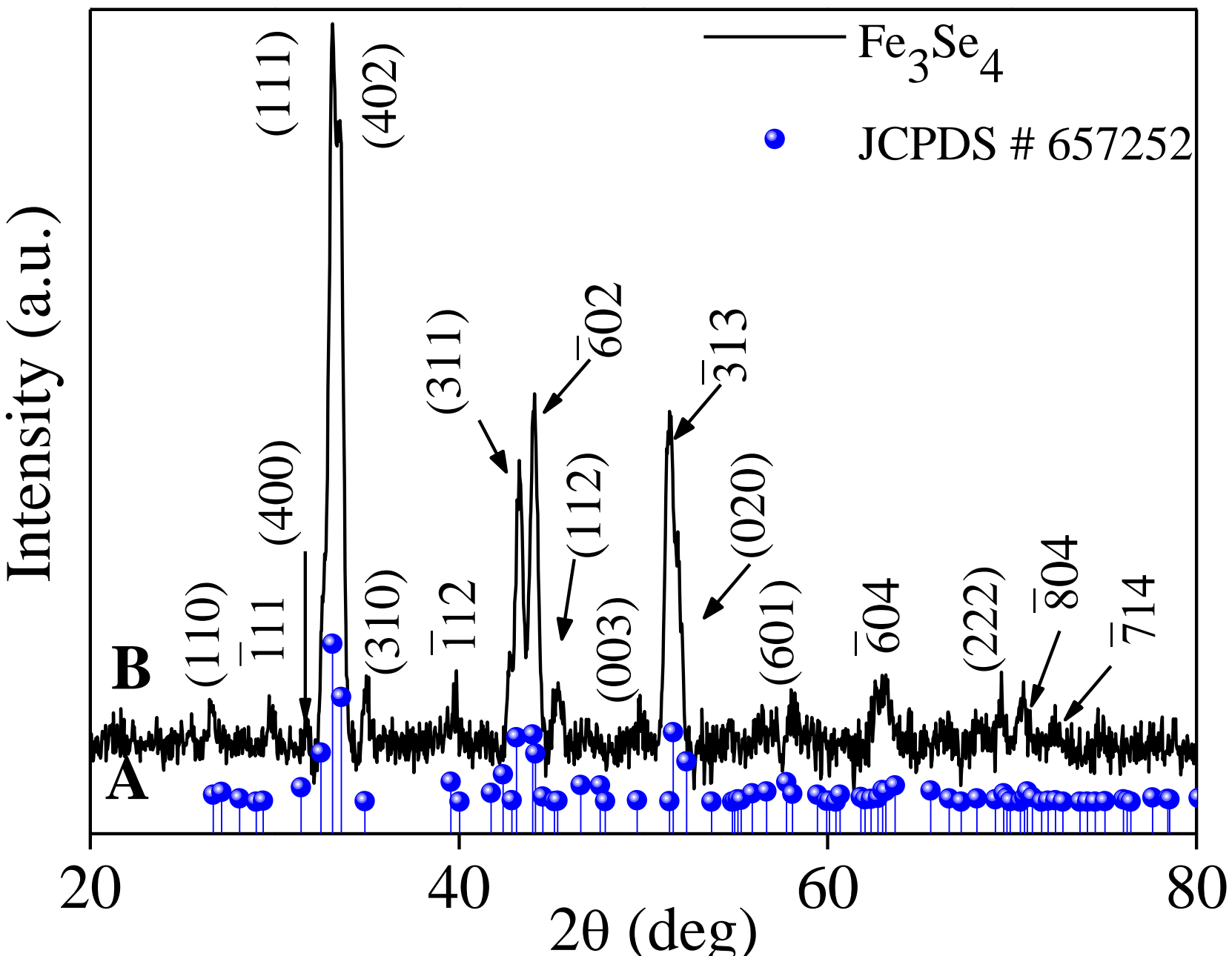

**Figure S1.** Powder XRD pattern of as synthesized $Fe_3Se_4$ nanoparticles.

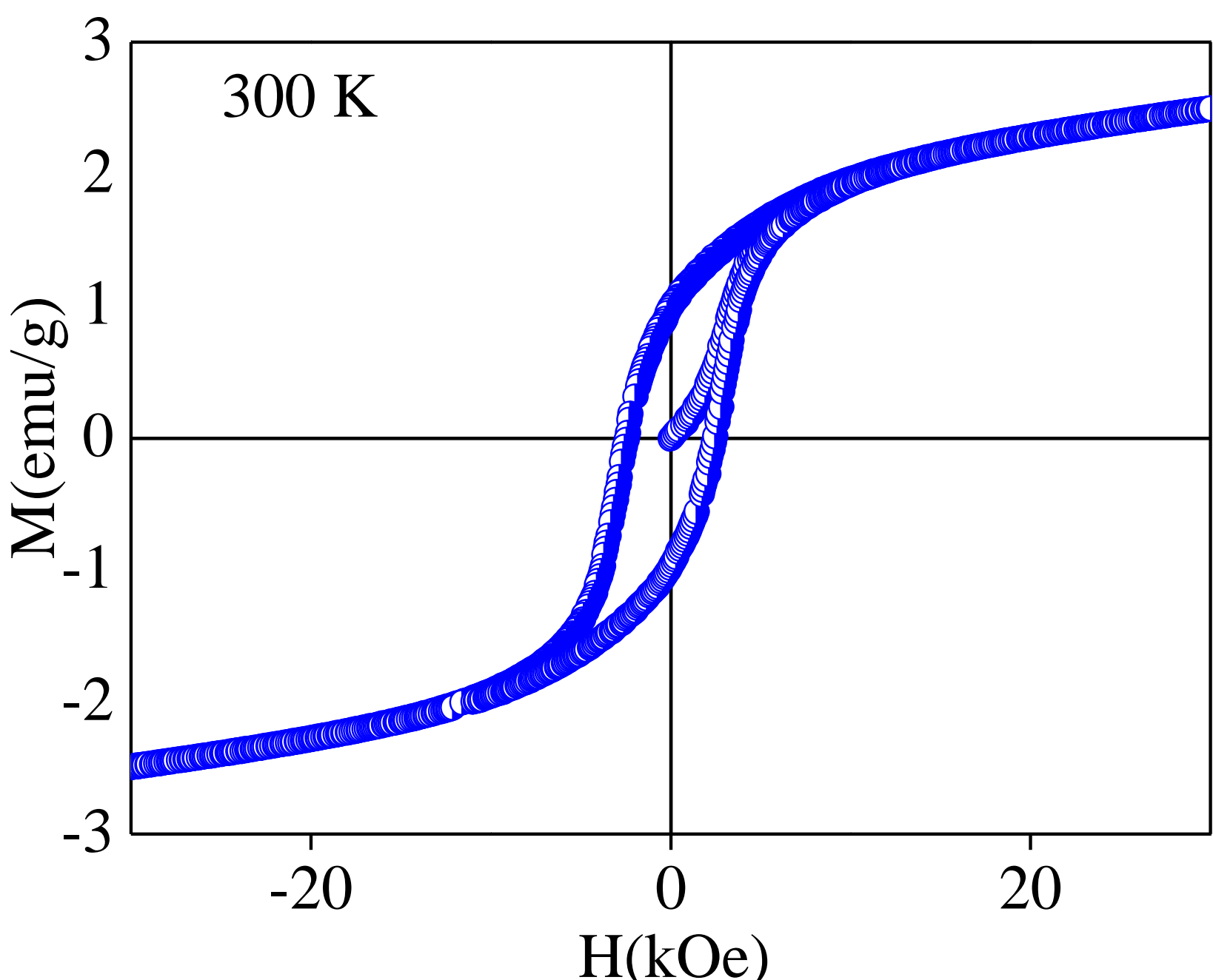

**Figure S2.** M-H hysteresis loop measurement of $Fe_3Se_4$ nanoparticles at 300 K.

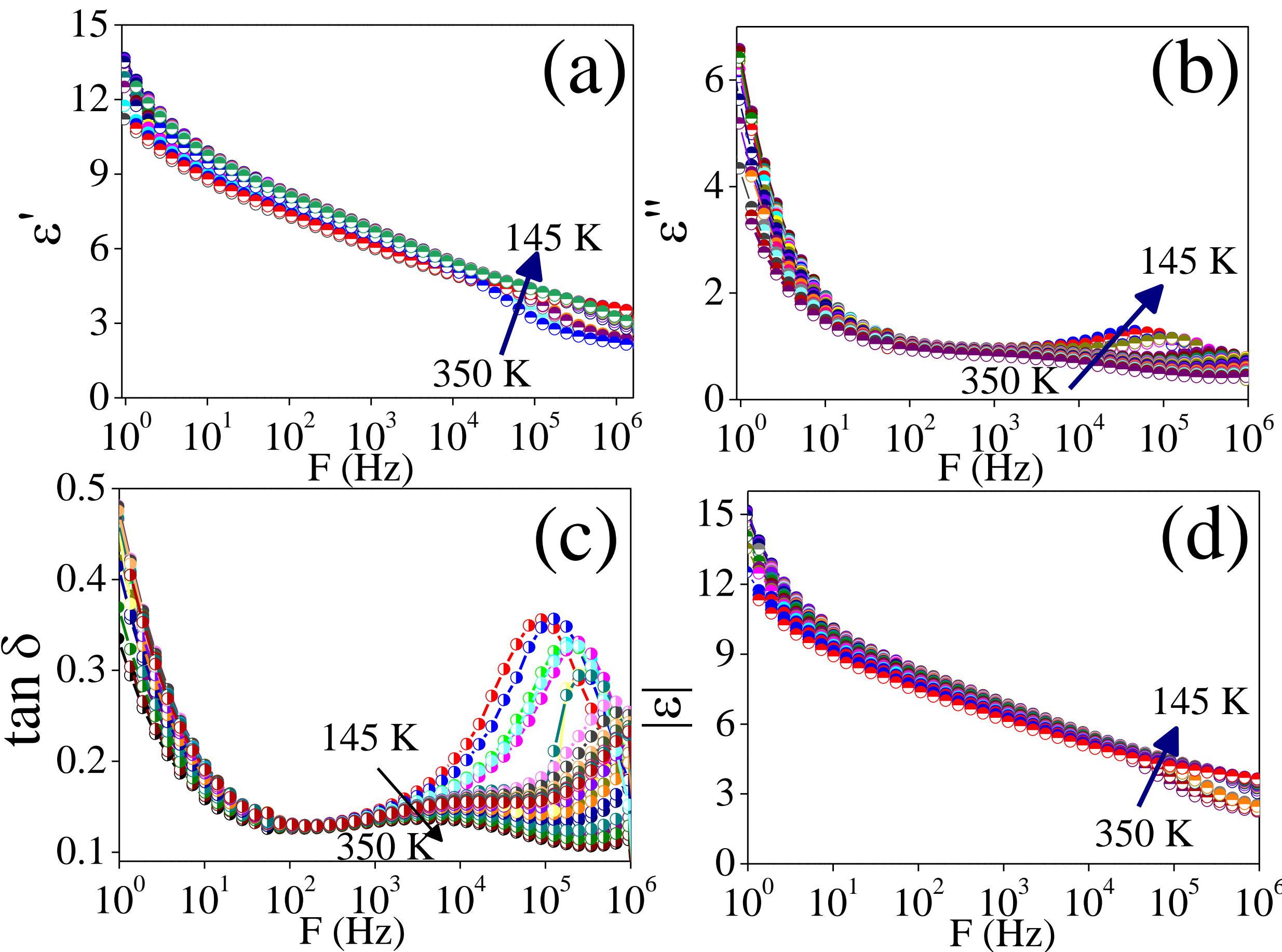

**Figure S3.** Frequency dependence of (a) real part of permittivity (b) out of phase part of permittivity (c) loss factor (d) absolute permittivity measured with ac 1 V rms value at a temperature range from 200 K to 350 K.

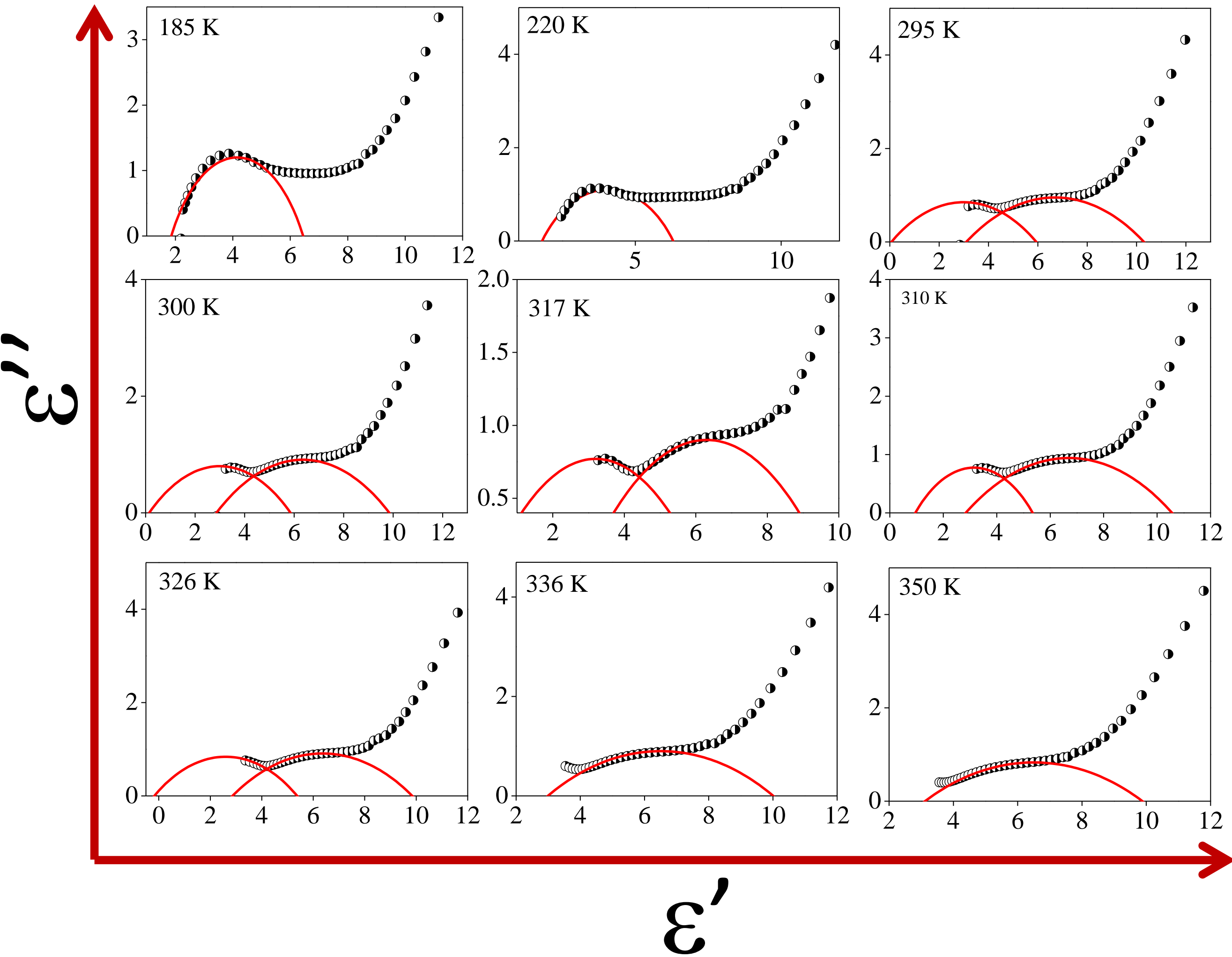

**Figure S4.** Cole-Cole plot of as synthesized $Fe_3Se_4$ nanoparticles (ε'' versus ε') measured by impedance spectroscopy at temperatures 185, 220, 295, 300, 317, 310, 326, 336 and 350 K respectively.

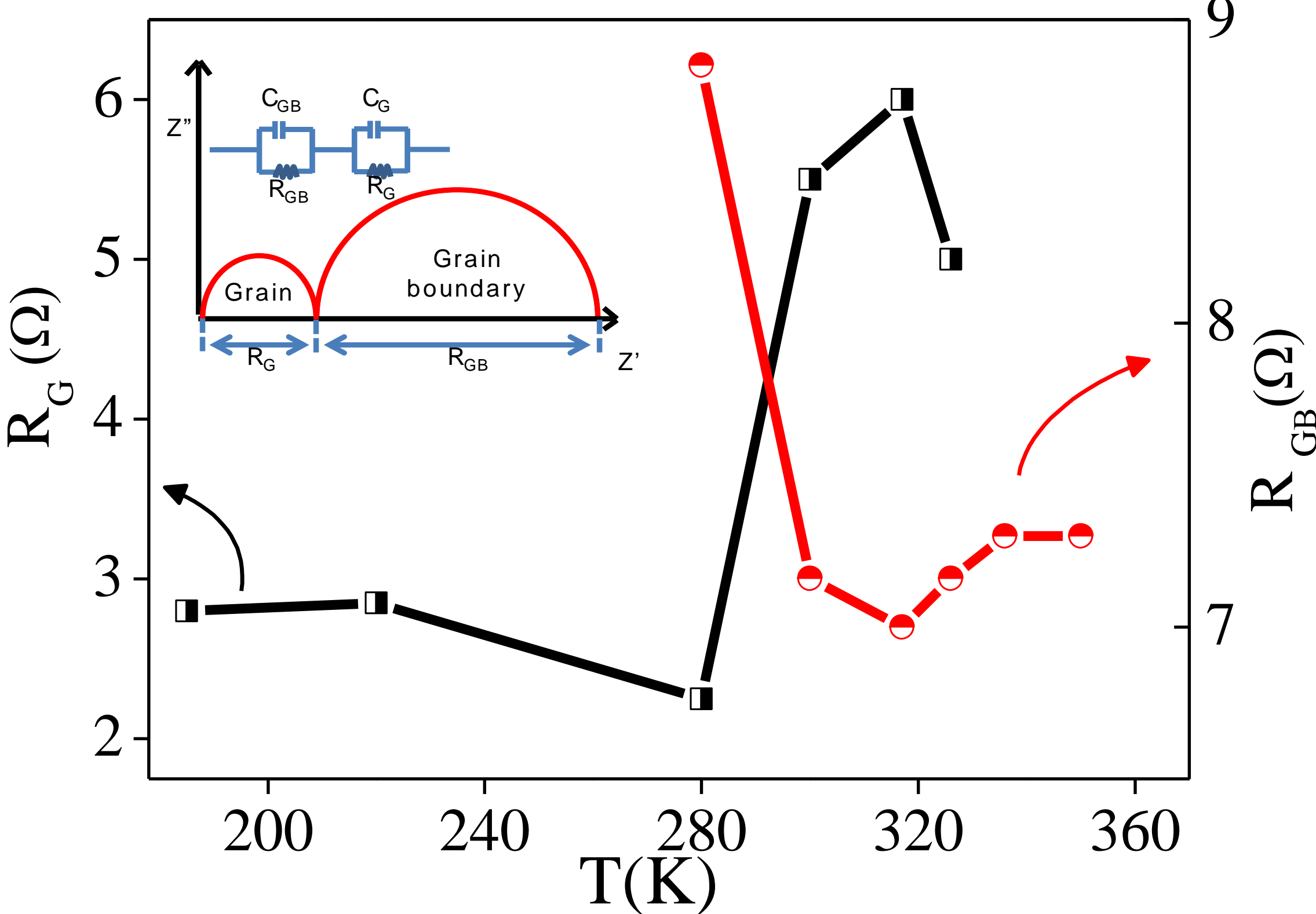

Figure S5 Temperature dependent of resistance contribution from grain and grain boundary extracted from Cole-Cole plot is plotted. Inset shows the circuit arrangements and parameters RG and RGB in a Cole-Cole plot.